\documentclass[aps,prl,twocolumn,superscriptaddress,showpacs]{revtex4}

\usepackage{times}
\usepackage{epsfig}
\usepackage{amsmath}
\usepackage{amssymb}

\begin{document}

\title{Multi-channel Kondo Models in non-Abelian Quantum Hall Droplets}

\author{Gregory A. Fiete}
\affiliation{Department of Physics, California Institute of Technology, MC 114-36, Pasadena, California 91125, USA}
\author{Waheb Bishara}
\affiliation{Department of Physics, California Institute of Technology, MC 114-36, Pasadena, California 91125, USA}
\author{Chetan Nayak}
\affiliation{Microsoft Research, Station Q, CNSI Building, University of California, Santa Barbara, California 93106, USA}
\affiliation{Department of Physics, University of California, Santa Barbara, California, 93106, USA}

\date{\today}

\begin{abstract}

We study the coupling between a quantum dot and the edge of a
non-Abelian fractional quantum Hall state which is spatially
separated from it by an integer quantum Hall state.
Near a resonance, the physics at energy scales below the level spacing of the edge states of the dot is governed by a $k$-channel Kondo model when the quantum Hall state is a Read-Rezayi state at filling fraction $\nu=2+k/(k+2)$ or its particle-hole
conjugate at $\nu=2+2/(k+2)$. The $k$-channel Kondo model is
channel isotropic even without fine tuning in the former state;
in the latter, it is generically channel anisotropic.
In the special case of $k=2$, our results provide a new venue, realized in a mesoscopic context, to distinguish between the Pfaffian and anti-Pfaffian states
at filling fraction $\nu=5/2$.

 \end{abstract}



\maketitle

Non-Abelian quantum Hall states have
sparked considerable interest recently because of their potential application
to topological quantum computing \cite{DasSarma:pt06}.
Though it is not known whether such states exist, it
is suspected that the observed plateaus at $\sigma_{xy} = \nu\,\frac{e^2}{h}$
with $\nu=5/2$ \cite{Willet:prl87}
and $\nu= 12/5$
\cite{Xia04} are due to non-Abelian quantum Hall states.
The evidence, thus far, comes primarily from numerical studies
\cite{Morf98,Rezayi00,Rezayi06} which found that the ground
states of small numbers of electrons had large overlap with
the Moore-Read Pfaffian wavefunction
\cite{Moore:npb91,Greiter92,Nayak96c}
and the particle-hole conjugate of the
$k=3$ Read-Rezayi (RR) wavefunction \cite{Read:prb99,Fiete_unpublished}
at $\nu=5/2$ and $\nu=12/5$ respectively, and from recent noise and tunneling measurements \cite{Dolev:cm08,Radu08}.
It has been argued
that these wavefunctions are representatives of
two universality classes which
exhibit non-Abelian quasi-particle statistics, which is a necessary ingredient for topological quantum computing \cite{Nayak08}.
Recently, further numerical studies \cite{Feiguin07b,Peterson08}
have bolstered the argument that
these states occur in the experiments of Refs
\cite{Willet:prl87,Xia04}.

Some theoretical proposals have been made to determine experimentally whether or not the $\nu=5/2$ state possesses the non-Abelian quasi-particle statistics of the Pfaffian
\cite{Fradkin98,DasSarma05,Stern:prl06,Bonderson:prl06}. 
While fabricating high-mobility samples of mesoscopic size to test these proposals presents a significant challenge, recent experiments on quantum point contacts at
$\nu=5/2$ give one reason to believe that such devices are within reach
\cite{Miller:nap07}. Experiments on such devices have recently shed
light, for the first time, on quasiparticle properties at $\nu=5/2$.
Shot noise \cite{Dolev:cm08} and non-linear
current-voltage characteristics \cite{Radu08} at quantum point contacts
at $\nu=5/2$ are both consistent with a quasi-particle charge of $e/4$,
as required by the Moore-Read Pfaffian state.

However, a wrinkle in the theoretical picture appeared recently
when it was realized that another state, the `anti-Pfaffian', is an
equally good candidate at $\nu=5/2$  \cite{Levin:prl07}.
The anti-Pfaffian is the conjugate of the Pfaffian under particle-hole
symmetry within a Landau level, which is an exact symmetry in the
limit of large magnetic field. This symmetry must be spontaneously
broken in order for one of these two degenerate ground states to occur;
the system sizes studied in numerics on the torus were simply too small to
observe anything other than the symmetric combination of the two
\cite{Rezayi00}. In the case of numerics on the sphere \cite{Morf98},
the finite geometry explicitly breaks the symmetry;
the anti-Pfaffian occurs at a different value
of the magnetic flux and was, consequently, missed.

The Pfaffian and anti-Pfaffian states
differ significantly in the nature of their edge excitations.
This leads to a difference in tunneling characteristics and
thermal transport along the edge \cite{Levin:prl07,Bishara:prb08}.
Both states possess charge $e/4$ quasi-particles, so the
existing noise experiments do not allow one to distinguish between them
\cite{Dolev:cm08}. Current-voltage characteristics at a point contact
can distinguish between the two states; measurements appear to be more consistent
with the anti-Pfaffian state although they cannot fully rule out the Pfaffian \cite{Radu08}.
Therefore, there is urgent need for further experiments to determine not only
whether the $\nu=5/2$ state is Abelian or non-Abelian, but to indicate
which non-Abelian state. In this Letter, we propose such experiments.
We generalize our discussion of the Pfaffian state to
cover the Read-Rezayi states, as well.
To date, a few theoretical proposals have been made to probe
the topological properties this state \cite{Ilan:prl08,Bonderson_2:prl06}.
Our proposal is largely orthogonal to the existing ones. 

\begin{figure}[b]
\includegraphics[width=.65\linewidth,clip=]{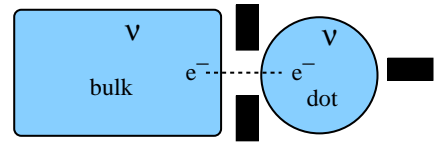}
 \caption{\label{fig:schematic} (color online) Schematic of our model.  Gates are shown in black. They may be used to form a point contact to pinch off the dot from the rest of the quantum Hall bulk.  The gate on the right of the figure may be used to shift the energy levels of the dot by changing its area $S$.  The bulk is assumed to be at filling fraction $\nu=2+k/(k+2)$ or $\nu=2+2/(k+2)$. The white region between the dot
and the bulk is assumed to be at $\nu=2$.
The charge on the dot may be measured capacitively \cite{Ashoori:prl92}.}
 \end{figure}

In this Letter, we study the set-up shown schematically in Fig.~\ref{fig:schematic}.
The bulk quantum Hall state on the left is assumed to be in a non-Abelian
fractional quantum Hall state at $\nu=2+k/(k+2)$ or $\nu=2+2/(k+2)$, as we describe below.
A quantum point contact may be used to pinch off a finite region of the quantum Hall fluid and form a quantum dot separated from the bulk by a tunneling barrier.  We assume that the lower two Landau levels are not pinched off and therefore do not backscatter at the point contact,
i.e. the barrier region is assumed to have $\nu=2$. 
For an infinite system, the edge modes of the quantum dot are gapless,
but for a finite system they acquire a discrete spectrum \cite{Fiete:prl07,Ilan:prl08}.
We focus on fluctuations of the quantum dot charge $Q \equiv e\langle \hat N_e \rangle$
near a degeneracy point in the energy, that is when the energy of a dot with $N_e$
electrons is equal to that of a dot with ${N_e}+1$ electrons:
$E({N_e},S,B)=E({N_e}+1,S,B)$. The energy of the dot depends on its area $S$,
 which may be altered by a gate potential shown in Fig.~\ref{fig:schematic}; and on
 the magnetic field $B$.  Adjusting either $S$ or $B$ may be used to achieve the desired degeneracy and also to slightly tune away from it. The charge of the dot can be measured capacitively \cite{Ashoori:prl92}.

We are interested in energy and temperature scales much less than the level-spacing
of the dot edge states.  Under this assumption, only two levels on the quantum dot
(the degenerate or nearly degenerate ones) are important for the physics.
In our formulation, these two levels will act as an effective, local spin-1/2 degree of freedom.
The crucial feature of the Read-Rezayi states is that the coupling of their edge states to this
effective spin degree of freedom via electron tunneling to the dot can be mapped to
the $k$-channel Kondo model. (We emphasize that our analysis applies to an
effective spin degree of freedom which accounts for the charge on the dot;
the Read-Rezayi states and their particle-hole conjugates \cite{Bishara_RR:prb08}
are spin polarized,
so there are no spin-flips in the quantum Hall edge states).
This allows us to exploit known results from the multi-channel Kondo models
\cite{Affleck:npb91}.

A detuning from degeneracy maps to the coupling of an external magnetic field to the spin in the $k$-channel Kondo model.  Thus, charge susceptibilities in our quantum dot set-up can be obtained from magnetic susceptibility in the Kondo model.  A remarkable feature of the scenario we discuss here is that the channel {\em isotropic} limit is automatically obtained for Read-Rezayi states {\em without any fine tuning}.  Again, this feature follows from the form of the coupling of the edge states to the quantum dot degrees of freedom.  On the other hand,
for their particle-hole conjugates \cite{Levin:prl07,Bishara_RR:prb08},
the generic case is channel anisotropic.
Since the channel isotropic and channel anisotropic Kondo models are very different,
one can exploit the thermodynamics of the multi-channel Kondo model applied to the charge susceptibility to distinguish the Pfaffian from the anti-Pfaffian. This is one of our central results.

{\it Pfaffian State.}
We begin with the case of a quantum dot coupled to a bulk quantum
Hall state in the Moore-Read Pfaffian state.
The Hamiltonian for our problem is $H=H_{\rm edge}+H_{\rm dot}+H_{\rm tun}$.
Here and henceforth, we ignore the $2$ filled lower Landau levels.  This is justified by the sequence of modes pinched off in a point contact \cite{Dolev:cm08}.
The edge theory for the Pfaffian state
is the product of a free, charged chiral bosonic sector
and a neutral Majorana sector. The edge Hamiltonian takes the form
\begin{equation}
\label{eq:edge}
H_{\rm edge}=  \int dx  \:\left({v_c}\frac{(k+2)/k}{4\pi}  (\partial_x \varphi(x))^2  \: + \: 
i v_n   \psi \partial_x \psi \right),
\end{equation}
Here, $k=2$ and $v_n<v_c$ is the velocity of the neutral mode(s).
The Hamiltonian of the dot describes a two level system which we can take to be
``empty" or ``occupied" (later to be mapped to ``up" or ``down")
\cite{Furusaki:prl02,Matveev:prb95}.  It thus has a fermionic character
and we label the fermionic annihilation (creation) operator for this state $d\, (d^\dagger)$.
Thus, $H_{\rm dot}=\epsilon_d\, d^\dagger d$,
where $\epsilon_d=0$ at the degeneracy point and $\epsilon_d \neq 0$ when one is tuned away from degeneracy.  The tunneling Hamiltonian is 
\begin{equation}
H_{\rm tun}=t(d^\dagger {\Psi_e}(0) + {\Psi_e^\dagger}(0) d)
+ V {d^\dagger}d \,{\Psi^\dagger_e}(0){\Psi_e}(0),
\end{equation}
where $t$ is the tunneling amplitude to the dot; $x=0$
is the location of the point contact; $V$ is the
Coulomb repulsion between the edge and the dot,
and $\Psi_e\, (\Psi_e^\dagger)$ is the annihilation (creation) operator for the
electron, $\Psi_e^\dagger=\psi e^{i2\varphi}$.
We use the convention ${\rm dim}[e^{i\alpha \varphi}]=\nu \frac{\alpha^2}{2}$,
so that ${\rm dim}[\Psi_e]=3/2$.

As a result of the scaling dim of $\Psi_e$, $t$ is naively irrelevant,
\begin{equation}
\label{eqn:RG-naive}
\frac{dt}{d\ell} = -\frac{1}{2}\,t + {\cal O}(tV) + {\cal O}(t^3) .
\end{equation}
However, for $V$ sufficiently large, $t$ flows to the
$2$-channel Kondo fixed point, not to $t=0$. (One might
guess this from the second term above, but we will
show this directly.)
To see this, we we apply a unitary transformation
$U=e^{2i {d^\dagger} d\, \varphi(0)}$ to 
$H$, which rotates $\varphi(0)$ out of the tunneling term.
$H$ now takes the form
\begin{multline}
\label{eqn:transformed-EK}
U H U^\dagger =
H_{\rm edge} +H_{\rm dot} +t\,\psi({d-d^\dagger} )\\
+\: \left(V-2{v_c}\right) {d^\dagger}d\, \partial_x \varphi(0).
\end{multline}
For $V-2{v_c}$, this is a purely quadratic theory
which can be solved exactly. Thus, $t$ is clearly relevant
in this limit; we will see below that it is actually relevant
over a range of values of $V$.
Note that only the Majorana combination
${d-d^\dagger} $ couples to the the quantum Hall edge.
This is precisely the same feature which leads to
non-Fermi liquid behavior in the two-channel
Kondo problem \cite{Emery:prb92}: the spectral function
$\text{Im}\left\langle{d^\dagger} d\right\rangle$
has both a $\delta$-function piece, coming from
${d^\dagger}+d$ and a Lorentzian piece coming
from $d-{d^\dagger} $. As we will see, the coupling
of a quantum dot to an anti-Pfaffian quantum Hall state
does not have this property.

To see the connection to the two-channel Kondo model,
it is useful to represent the two-level system on
the dot by a spin: $S^\dagger = d^\dagger, S^-=d$,
and $S^z=d^\dagger d -1/2$ (up to Klein factors we have suppressed). 
Then, we apply a unitary transformation
$U=e^{i\alpha S^z \varphi(0)}$
to $H$ as before, but now we take $\alpha=2-\sqrt{2}$, i.e.
rather than fully rotate $\varphi(0)$ out of the tunneling term,
we partially rotate it. $H$ now takes the form:
\begin{multline}
\label{eqn:transformed}
U H U^\dagger =
H_{\rm edge} + \epsilon_d S^z + \left(V-
{v_c}\alpha \right) S^z \partial_x \varphi(0)\\
+t(\psi^\dagger e^{-i \sqrt{2}\varphi(0)}S^\dagger+
\psi e^{i \sqrt{2}\varphi(0)}{S^-}).
\end{multline}
We now compare this to the Hamiltonian for the
Kondo model:
\begin{equation}
\label{eqn:Kondo-coupling}
H_{\rm imp}= \lambda_\perp ( {J^+}(0)S^{-} + {J^-}(0)S^{+} )
+ \lambda_z {J^z}(0)S^{z} + h {S^z},
\end{equation}
where ${\vec S}$ is the impurity spin; ${\vec J}(0)$ is 
conduction electron spin density at the impurity site;
$\lambda_\perp$, $\lambda_z$ are the exchange
couplings which are not assumed to be equal;
and $h$ is the magnetic field.
The impurity spin only interacts with conduction electrons
in the $s$-wave channel about the origin. Retaining only this
channel, we have a chiral one-dimensional problem in which
the impurity is at the origin and the incoming and outgoing
modes are right-moving modes at $x<0$ and $x>0$, respectively.
Affleck and Ludwig observed 
\cite{Affleck:npb90,Affleck_2:npb91} that the Hamiltonian of
the conduction electrons $H_{\rm cond}$ in the $k$-channel
Kondo model admits a {\it conformal decomposition},
$H_{\rm cond}=H_{U(1)}+H_{SU(2)_k}+H_{SU(k)_2}$.
This decomposition reflects the SU(2)$_k$
Kac-Moody algebra satisfied by the spin density ${J^a}$, 
which we now exploit in the $k=2$ case and,
later, in the general case. For $k=2$,
we can express the ${J^a}$ in terms of a Majorana
fermion, $\psi$, and a free boson $\varphi$:
\begin{equation}
\label{eqn:current-rep}
J^\dagger=\sqrt{2}\psi e^{i \sqrt{2} \varphi},{\hskip 0.2 cm}
J^-=\sqrt{2}\psi e^{-i \sqrt{2} \varphi},{\hskip 0.2 cm}
J^z=\sqrt{2} \partial_x \varphi.
\end{equation}
The operators $\psi$ and $\varphi$ have a complicated,
non-local relation to the original conduction electron operators,
but they have the virtue of satisfying the SU(2)$_2$
Kac-Moody algebra via (\ref{eqn:current-rep}).

Substituting (\ref{eqn:current-rep}) into (\ref{eqn:Kondo-coupling}),
we see that our problem (\ref{eqn:transformed})
maps onto the $2$-channel Kondo model
with anisotropic exchange if we identify ${\lambda_\perp}=t$,
$\sqrt 2{\lambda_z}=V-(2-\sqrt{2}){v_c}$, and $h=\epsilon_d$.
For ${\lambda_z}<0$,
the Kondo model is ferromagnetic. In the ferromagnetic
Kondo model, the coupling to the impurity is irrelevant, as we naively
expected above (\ref{eqn:RG-naive}). However, when $V$
is sufficiently large, ${\lambda_z}>0$,
corresponding to the antiferromagnetic Kondo model.
In this case, the Hamiltonian is controlled in the infrared
by the exchange and channel isotropic antiferromagnetic
spin-1/2 $2$-channel Kondo fixed point \cite{Affleck:prb92}.
This fixed point is characterized by non-Fermi liquid
correlations, including anomalous exponents for
the temperature dependence of the impurity contribution
to the specific heat and spin susceptibility and the magnetic field
dependence of the zero-temperature magnetization. The latter
two translate to the charge susceptibility and charge of the
quantum dot \cite{Affleck:npb91}:
\begin{equation}
\chi_\text{charge} \propto \ln\left({T_K}/T\right)\: , \hskip 0.2 cm
{\Delta Q} \propto {V_G} \ln\left({k_B}{T_K}/{e^*}{V_G}\right),
\end{equation}
where the Kondo temperature depends on non-universal values $v_n,t$ and is given by ${T_K}\propto\exp(-c_1{v_n}/t)$ 
with $c_1>0$.
Here, $\Delta Q = Q-e({N_e}+\frac{1}{2})$ is the charge on
the dot measured relative to the average electron number
at the degeneracy point of the energy. In the case $k=2$,
which corresponds to the Pfaffian state, possibly realized
at $\nu=5/2$, there are logarithmic corrections:
Ordinarily, fine tuning would be required
to realize channel isotropy in the Kondo model
\cite{Oreg:prl03} but, as we have seen,
the coupling between a quantum dot and the edge
of a Pfaffian quantum Hall state automatically
realizes the channel isotropic $2$-channel Kondo model.

{\it Anti-Pfaffian State.}
We now turn to the coupling of the anti-Pfaffian state to the
two degenerate levels of the quantum dot. The edge
theory of the anti-Pfaffian state is \cite{Levin:prl07}: 
\begin{equation}
{\cal L}_{\overline{\rm Pf}}
= \frac{2}{4 \pi} {\partial_x}{\phi_\rho}
({\partial_t}-v_{\rho}{\partial_x}){\phi_\rho}
+ i{\psi_a}(-{\partial_t}-
v_{\sigma}{\partial_x}){\psi_a}.
\end{equation}
There is a charged boson $\phi_\rho$ and three
counter-propagating Majorana fermions $\psi_a$,
$a=1,2,3$. There are three different dimension-$3/2$
electron operators, ${\psi_a} e^{2i\phi_\rho}$. The combination
$({\psi_1} - i {\psi_2})e^{2i\phi_\rho}$ is inherited from the
electron operator of the $\nu=1$ integer quantum Hall state
in which a Pfaffian state of holes forms. Thus, we expect it
to have the largest tunneling amplitude. The other two
electron operators are complicated charge-$e$ combinations of the 
$\nu=1$ electron operator and the electron operator of the
Pfaffian state of holes. The tunneling Hamiltonian is (the repeated index $a$ is
summed over):
\begin{equation}
H_{\rm tun}= \left(
{t_a}\psi_a e^{-2i\phi_\rho} d^\dagger + \text{h.c.}\right)
+ V {d^\dagger} d {\partial_x} {\phi_\rho}
\end{equation}
Performing a unitary transformation as before to rotate out
the $\phi_\rho$ dependence of the first term, we obtain
$U H {U^\dagger} = H_\text{edge} + H_\text{dot} +
{\tilde H}_\text{tun}$ where 
\begin{multline}
\label{eqn:anti-Pfaff-Toulouse}
{\tilde H}_\text{tun} =
\left(
{t_a}\psi_a d^\dagger + \text{h.c.}\right)
+ \left(V-2{v_c}\right) {d^\dagger} d {\partial_x} {\phi_\rho}\\
= i{\chi^{}_1}\left( {\lambda_1} ({d^\dagger}-d)/i  + {\lambda'_1}({d^\dagger}+d)\right)\\
+ i{\lambda_2}{\chi^{}_2}({d^\dagger}+d)
\:+ \left(V-2{v_c}\right) {d^\dagger} d \, {\partial_x} {\phi_\rho}
\end{multline}
where ${\chi_1}={u_a}{\psi_a}/\sqrt{u^2}$, ${\chi_2}={w_a}{\psi_a}/\sqrt{w^2}$,
${u_a}=\text{Re}\,{t_a}$, ${v_a}=\text{Im}\,{t_a}$, ${w_a}={v_a}-{u_a}(u\cdot v/{u^2})$,
${\lambda_1}=\sqrt{u^2}$, ${\lambda_2}=\sqrt{w^2}$, ${\lambda'_1}=u\cdot v/\sqrt{u^2}$,
and $\{{\chi_1},{\chi_2}\}=0$.
Note that, for generic $t_a$s, both $({d^\dagger}-d)$ and $({d^\dagger}+d)/i$
couple to the edge modes,
as in the {\it one-channel} Kondo model.
This is in contrast to the Pfaffian case, in which only $({d^\dagger}-d)$
couples to the edge modes, as in the two-channel Kondo model.
At the Toulouse point, the {\it one-channel} Kondo model can be mapped
to a form similar to (\ref{eqn:anti-Pfaff-Toulouse}) with $V=2{v_c}$.
The charge susceptibility and charge of the
quantum dot have the temperature and voltage dependence
characteristic of a Fermi liquid:
\begin{equation}
\chi_\text{charge} \propto \text{const.}\:\:, \hskip 0.5 cm
{\Delta Q} \propto {V_G}.
\end{equation}
Consequently, measurements of the behavior of the dot would
distinguish the Pfaffian and anti-Pfaffian states.

{\it Read-Rezayi State.}
Now we analyze the Read-Rezayi state
with filling $k/(k+2)$, generalizing our
discussion above of the Pfaffian state,
which is the $k=2$ case.
The $k=1$ case is the Laughlin $\nu=1/3$ state,
while the $k=3$ case is the $3/5$ Read-Rezayi state.
The edge Hamiltonian takes the form $H_{\rm edge}=H_c+H_{Z_k}$
with $H_c$ the same as the first term in Eq. \eqref{eq:edge}.
$H_{Z_k}$ can be written as a gauged SU(2)$_k$ WZW model
in which the U(1) subgroup has been gauged,
thereby realizing an SU(2)$_k$/U(1) coset with central charge
$c=\frac{3k}{k+2}-1=\frac{2k-2}{k+2}$,
but we will not need this representation here.
The electron operator now takes the form:
$\Psi_e^\dagger=\psi_1 e^{i\frac{k+2}{k}\varphi}$ where $\psi_1$ is a parafermion field \cite{CFT}.  Since ${\rm dim}[\psi_1]=1-1/k$, ${\rm dim}[\Psi_e]=3/2$.

As before, we apply a unitary transformation
$U=e^{i\alpha S^z \varphi(0)}$ to $H$, which now takes the form
\begin{multline}
\label{eqn:transformed-RR}
U H U^\dagger =
H_{\rm edge} +H_{\rm dot} + \left(V-
{v_c}\alpha \right) S^z \partial_x \varphi(0)\\
+t(\psi_1^\dagger e^{-i\tilde \alpha\varphi(0)}S^\dagger+
\psi_1e^{i\tilde \alpha\varphi(0)}{S^-})
\end{multline}
where $\tilde \alpha\equiv \frac{k+2}{k}-\alpha$.
The choice $\alpha^* = \frac{k+2}{k}-\sqrt{\frac{k+2}{k}}\sqrt{\frac{2}{k}}$
makes the connection to the $k$-channel Kondo problem explicit
because the SU(2)$_k$ current operators
can be represented in terms of the $\mathbb{Z}_k$ parafermions:
$J^\dagger=\sqrt{k}\psi_1 e^{i \beta \varphi},J^-=\sqrt{k}\psi_1^\dagger e^{-i \beta \varphi},
J^z={k \over 2}\beta \partial_x \varphi$
where $\beta=\sqrt{2(k+2)}/k$.
Substituting these expressions into 
(\ref{eqn:Kondo-coupling}) we see that our problem
(\ref{eqn:transformed-RR})
is equivalent to the $k$-channel Kondo problem
if we identify ${\lambda_\perp}=t$,
$\beta {\lambda_z}=V-{v_c}\alpha^* $, and $h={\epsilon_d}$.
For $V>{v_c}\alpha^* $, this is the antiferromagnetic Kondo
problem, which has an intermediate coupling fixed point.
Thus, we see that the Read-Rezayi states offer a novel scenario to
realize the non-Fermi liquid behavior of the $k$-channel Kondo model,
$\chi_\text{charge} \propto T^{-(k-2)/(k+2)}, {\Delta Q} \propto {V_G}^{2/k}$
which would otherwise require an incredible amount of fine-tuning
for $k\geq 3$. Moreover, observing the predicted non-Fermi liquid behavior
would provide strong evidence that the quantum Hall state
is of the Read-Rezayi type.  It can also be shown that the particle-hole conjugate of the Read-Rezayi state at $\nu=2+2/(k+2)$ generalizes the result obtained above for the anti-Pfaffian:  the Kondo model realized is not channel isotropic \cite{Fiete_unpublished}.

{\it Tunneling through a Quantum Dot.}
We now consider the situation of bulk $\nu=2+k/(k+2)$
quantum Hall states on either side of a quantum dot. By the arguments
above, if the two tunneling amplitudes $t_L$ and $t_R$ are equal\cite{Kane:prb92}
then this model can be mapped onto the $2k$-channel Kondo model.
Consequently, the conductance through the dot is finite at zero
temperature,
$G = G_0 - a T^{2/(k+2)}$
where $G_0$, $a$ are constants \cite{Ludwig:prl91,Affleck93}. However, if ${t_L}\neq{t_R}$,
the smaller one scales to zero in the infrared and the conductance
through the dot vanishes. The temperature and voltage dependence
of the charge on the dot is governed by the $k$-channel Kondo model,
as before.

In summary, we have shown that a quantum dot coupled via tunneling to
a Pfaffian quantum Hall state realizes the channel isotropic $2$-channel Kondo model
while a quantum dot coupled to a Read-Rezayi state of filling factor
$\nu=2+k/(k+2)$ leads to a channel isotropic $k$-channel Kondo problem, both without any fine tuning of parameters.  These systems will thus exhibit all the known non-Fermi liquid properties in their thermodynamics, expressed through the charge on the dot, which may be measured capacitively. Because the coupling of a quantum dot to an anti-Pfaffian
state generically exhibits Fermi liquid properties our results may be used to distinguish between the two leading candidate states for $\nu=5/2$, the Pfaffian and the anti-Pfaffian.

We acknowledge helpful discussions with Eddy Ardonne, Jim Eisenstein, Lukasz Fidkowski,
Andreas Ludwig, and Gil Refael.  GAF would especially like to thank E. Ardonne for many discussions on conformal field theory.  GAF was supported by the Lee A. DuBridge Foundation.
\vskip -0.2 cm


\end{document}